\documentclass[conference]{IEEEtran}
\IEEEoverridecommandlockouts
\usepackage[style=ieee, backend=biber]{biblatex}
\usepackage{amsmath,amssymb,amsfonts}
\usepackage{graphicx}
\usepackage{textcomp}
\usepackage{xcolor}

\usepackage{multirow}
\usepackage{booktabs}
\usepackage{diagbox}

\usepackage{algorithm}
\usepackage[noend]{algpseudocode}

\def\BibTeX{{\rm B\kern-.05em{\sc i\kern-.025em b}\kern-.08em
    T\kern-.1667em\lower.7ex\hbox{E}\kern-.125emX}}
\begin{document}

\title{Three-Input Ciphertext Multiplication for Homomorphic Encryption\\
}

\author{\IEEEauthorblockN{Sajjad Akherati, Yok Jye Tang, and Xinmiao Zhang}
		\IEEEauthorblockA{
The Ohio State University, OH 43210, U.S.\\
Emails: \{akherati.1, tang.1121, zhang.8952\}@osu.edu}
 }

\maketitle

\begin{abstract} Homomorphic encryption (HE) allows computations to be directly carried out on ciphertexts and is essential to privacy-preserving computing, such as neural network inference, medical diagnosis, and financial data analysis. Only addition and 2-input multiplication are defined over ciphertexts in popular HE schemes. However, many HE applications involve non-linear functions and they need to be approximated using high-order polynomials to maintain precision. To reduce the complexity of these computations, this paper proposes 3-input ciphertext multiplication. One extra evaluation key is introduced to carry out the relinearization step of ciphertext multiplication, and new formulas are proposed to combine computations and share intermediate results. Compared to using two consecutive 2-input multiplications, computing the product of three ciphertexts utilizing the proposed scheme leads to almost a half of the latency, 29\% smaller silicon area, and lower noise without scarifying the throughput. 
\end{abstract}


\section{Introduction}
Homomorphic encryption (HE) allows computations to be directly carried out on encrypted data. It is indispensable to privacy-preserving machine learning, medical diagnosis \parencite{medical_diagnosis1, medical-diagnosis-3}, financial data analysis \parencite{finance}, and many other applications. 

For popular HE schemes \parencite{BGV, BV, FV, CKKS}, a ciphertext consists of two polynomials in the ring $\mathcal{R}_Q=\mathbb Z_Q/(x^N+1)$, i.e. $ct=(c_0(x),c_1(x))\in \mathcal{R}_Q^2$. $N$ is in the order of thousands and the modulus $Q$  has hundreds of bits. Ciphertext additions are coefficient-wise modulo $Q$ additions. Ciphtertext multiplication consists of two major steps: polynomial multiplication and relinearization. The complexity of long polynomial multiplications can be reduced by decomposing the polynomials and integrating the modular reduction by $x^N+1$ into the decomposed multiplication \cite{PolyMultSiPS, PolyMultJourn} as well as Number Theoretic Transform (NTT) \parencite{ParhiNTT, HanhoNTT}. Integer modular multiplications can be simplified utilizing Montgomery multiplication, Barrett reduction, and Karatsuba decomposition \parencite{ZhangMultSiPS, ZhangMult}. Its complexity can be reduced substantially by utilizing the residue number system (RNS) moduli \parencite{RNSCKKS, FV-RNS}, in which an integer $q \mod Q$ is represented by its modular reduction results by the co-prime factors of $Q$. 

%Specifically, in the RNS-CKKS scheme \parencite{RNSCKKS}, the modulus $Q=q_0q_1\cdots q_L$. The parameter $L$ denotes the depth of evaluations, which  means $L$ ciphertext multiplications can be evaluated before performing the expensive Bootstrapping process \parencite{Bootstrapping1, Bootstrapping4}.  As a result, an integer $a \mod Q$, denoted by $[a]_Q$, is represented as $[a]_{q_0},\cdots,[a]_{q_{L-1}}$ and the multiplication/additions are carried out on $[a]_{q_j}$ ($0\leq j< L$), which are shorter integers. 

%\bibitem{ZhangMultSiPS} Z. Huai, K. K. Parhi, and X. Zhang, ``Efficient architecture for long integer modular multiplication over Solinas prime,” {\it Proc. of IEEE Workshop on Signal Processing Syst.}, pp. 146-151, Oct. 2021.

%\bibitem {ZhangMult} Z. Huai, J. Zhou, and X. Zhang, ``Efficient hardware implementation architectures for long integer modular multiplication over general Solinas prime,” {\it Springer Journ. of Signal Processing Systems}, vol. 94, pp. 1067-1082, Aug. 2022.

Despite all the improvements, ciphertext multiplication has high complexity and it is limited to 2-input in existing literature. On the other hand, many applications involve the multiplication of multiple ciphertexts, such as higher-order approximations of the ReLU function in neural networks \parencite{ConvFHE} and decision trees for medical diagnosis \parencite{medical-diagnosis-3}. Multiplying two ciphertexts at a time leads to long latency.   

%ciphertext multiplication is one of the most complex HE operations which involve three steps, including polynomial multiplications between the ciphertext's elements, relinearization and rescaling. The relinearization step prevents the ciphertext's size to grow exponentially and rescaling avoid the noise increase. 

This paper proposes a 3-input ciphertext multiplication scheme for the popular RNS-CKKS HE \cite{RNSCKKS}. An extra evaluation key is introduced to handle the relinearization resulted from the third input ciphertext, and the multiplications of the input polynomials are reformulated to reduce the complexity. Besides, the proposed relinearization is further reformulated to reduce the expensive modulus switching  needed and the extra evaluation key is chosen to reduce the number of polynomial multiplications. Mathematical analysis shows that our proposed scheme leads to lower noise compared to computing the product of multiple ciphertexts using prior 2-input multiplications. Hardware implementation architecture is also developed in this paper. Compared to carrying out two 2-input multiplications one after another, the proposed 3-input ciphertext multiplication leads to around 50\% shorter latency and 29\% smaller area for an example case that $Q$ is decomposed into three factors of 30 bits. 

\vspace{-0.5em}

%This method can be extended to handle the multiplication of more than three ciphertexts, achieving even greater complexity reduction.

\begin{figure*}[t]
		\centering
		\includegraphics[scale=0.7]{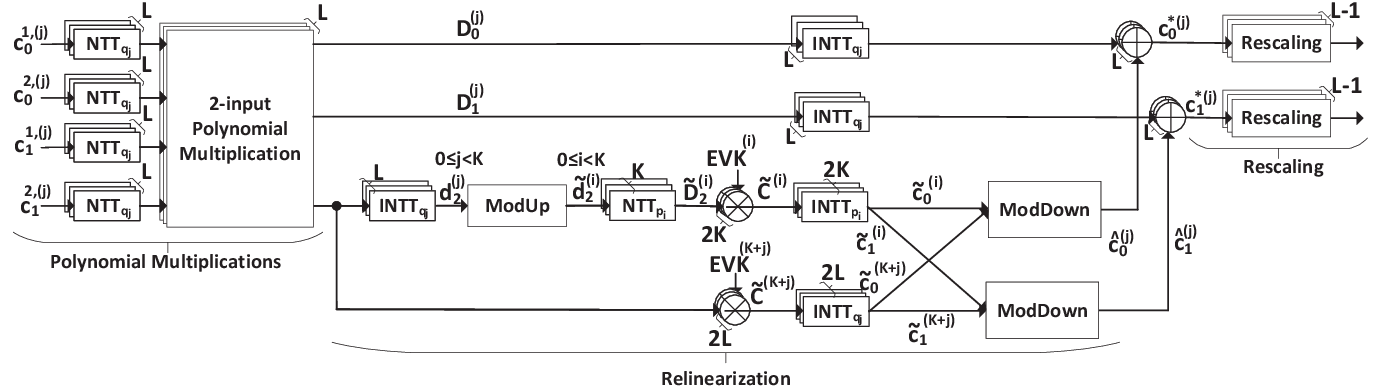}
		\caption{The block diagram for 2-input ciphertext multiplication in the RNS-CKSS scheme \cite{RNSCKKS}} \label{fig: 2-ct_mult}
\end{figure*}

\section{RNS-CKKS Ciphertext Multiplication}
In this paper, the vector and polynomial representations are utilized interchangeably. In the CKKS HE scheme, each user chooses a secret key, $sk=(1,s)$, where $s$ is a length-$N$ random signed binary vector with the weight decided by the security parameter. The public key is set to $pk=(b,a)\in \mathcal R^2_Q$, where $a$ is randomly generated, $b=(-as+e) \mod Q$, and $e$ is a random Gaussian vector of low norm. Utilizing a random signed binary vector, $v$, and two random Gaussian vectors of low norm, $e_0$ and $e_1$, a message $m$ is encrypted to ciphertext $ct=(c_0,c_1)\in \mathcal R_Q^2=v\cdot pk+(m+e_0,e_1)\mod Q$. Given a ciphertext $ct=(c_0,c_1)$, the decrypted message is $\lceil(c_0+c_1s)\mod Q\rfloor$. Besides, there is an evaluation key used for the relinearization step of ciphertext multiplication $evk = (evk_0,evk_1)\in \mathcal R_{PQ}^2=(-s\cdot evk_1+e'+Ps^2, evk_1)\mod (PQ)$, where $P$ is another large integer that has a similar number of bits as $Q$, $evk_1$ is sampled from $\mathcal R_{PQ}$, and $e'$ is a random vector with low norm.

In the RNS representation, assume that $Q=q_0q_1\cdots q_{L-1}$, and $q_j$ $(0\leq j<L)$ are co-prime to each other. Then an integer, $r$, mod $Q$ can be uniquely represented by $r_j=r \mod q_j$, and the computations over $r$ can be mapped to those over $r_j$ instead. Since $r_j$ are much smaller, the computation complexity is greatly reduced. 

In the RNS-CKKS scheme \cite{RNSCKKS}, each ciphertext is represented by $L$ elements as $ct=[c^{(0)},c^{(1)},\cdots,c^{(L)}]$, where $c^{(j)}=(c_0^{(j)},c_1^{(j)})\in \mathcal R_{q_j}^2$ ($0\leq j\leq L$). The multiplication of two ciphertexts, $ct^1=[c^{1,(0)},c^{1,(1)},\cdots,c^{1,(L)}]$ and $ct^2=[c^{2,(0)},c^{2,(1)},\cdots,c^{2,(L)}]$, is described in Fig. \ref{fig: 2-ct_mult}. It consists of two major steps: polynomial multiplication and relinearization. The polynomial multiplication step computes $d^{(j)}_0=c_0^{1,(j)}c_0^{2,(j)}$, $d^{(j)}_1=c_0^{1,(j)}c_1^{2,(j)}+c_1^{1,(j)}c_0^{2,(j)}$, and $d^{(j)}_2=c_1^{1,(j)}c_1^{2,(j)}$. These
three polynomials are reduced to two polynomials through the relinearization process in order to keep ciphertext format consistent for succeeding computations. 

Denote the product of $ct_0$ and $ct_1$ by $ct^*=(c^*_0,c^*_1)$. The relinearization step computes $ct^*_0=(d_0+\lceil d_2\cdot evk_0/P\rfloor)\mod Q$ and $ct^*_1=(d_1+\lceil d_2\cdot evk_1/P\rfloor)\mod Q$. When RNS representation is adopted, the moduli of $d_2$ need to be increased by a ModUp procedure before it is multiplied to $evk$, since $evk$ has a larger modulus. Besides, after the multiplication, the modulus of the product needs to be reduced to $Q$ by utilizing a ModDown process \cite{RNSCKKS,Combined}. Assume that $P=p_0p_1\cdots p_{K-1}$, and $p_i$ $(0\leq i<K)$ are co-prime to each other. Both the ModUp and ModDown mainly consist of a basis conversion process, which is simplified by the fast basis conversion formula \cite{FBC} as $\sum_{j=0}^{L-1}(a^{j}\hat{q}_j^{-1} \mod {q_j})\cdot\hat{q_j} \mod {p_i}$. Here $\hat{q_j} = \prod_{j'=0, j'\neq j}^{L-1}q_{j'}$ ($0\!\!\leq j\!\!<L$). The CKKS scheme also has a re-scaling process at the end, which computes $q_{L-1}^{-1}(c_0^{*(j)}-c_0^{*(L-1)}) \mod q_{j}$. The purpose of dividing $q_{L-1}$ is to reduce the noise. 

NTT can be applied to reduce the complexity of polynomial multiplications as shown in Fig. \ref{fig: 2-ct_mult}. In this figure, capital alphabets are used to denote the corresponding polynomials in the transformed domain. For example, $D_0^{(j)}$ denotes NTT of $d_0^{(j)}$. In Fig. \ref{fig: 2-ct_mult}, NTT is first applied to each input polynomial. Then the polynomial multiplications are implemented as coefficient-wise multiplications. Since ModUp is nonlinear, $D_2^{(j)}$ needs to go through inverse (I-)NTT before ModUp is applied. Still the multiplication with $evk$ can be greatly simplified in the transformed domain. Hence, the NTT of $evk$ is pre-computed and NTT is applied again on the ModUp output. Similarly, ModDown is nonlinear and INTT is applied at its input.

\section{Three-input Ciphertext Multiplication}
Ciphertext multiplication is limited to 2-input in existing literature. Multi-input multiplication will greatly speed up the evaluation of high-order polynomials and decision trees that are used in numerous neural networks and medical diagnosis applications. This section proposes a 3-input ciphertext multiplication by introducing an extra evaluation key. To reduce the hardware complexity, the involved computations are reformulated to combine calculations and share intermediate results. Besides, noise analysis is provided to show that the proposed scheme leads to lower noise level in the product ciphertext compared to using 2-input multiplications twice to compute the product of three inputs.

\subsection{Proposed 3-input ciphertext multiplication}
Consider the multiplication of three ciphertexts in the CKKS scheme, $ct^1=(c^1_0,c^1_1)$, $ct^2=(c^2_0,c^2_1)$, and $ct^3=(c^3_0,c^3_1)$. They are decrypted to $c^1_0+c^1_1s$, $c^2_0+c^2_1s$, and $c^3_0+c^3_1s$, respectively, and their product is $(c^0_0+c^0_1s)(c^1_0+c^1_1s)(c^2_0+c^2_1s)=d_0+d_1s+d_2s^2+d_3s^3$, where 
\begin{equation}\label{eq: mult3}
\begin{split}
&d_0=c^1_0c^2_0c^3_0,\ \  d_1=c^1_1c^2_0c^3_0+c^1_0c^2_1c^3_0+c^1_0c^2_0c^3_1\\
&d_2=c^1_1c^2_1c^3_0+c^1_1c^2_0c^3_1+c^1_0c^2_1c^3_1,\ \  d_3=c^1_1c^2_1c^3_1.
\end{split}
\end{equation}
The first two terms in the above formula contribute to the $c^*_0$ and $c^*_1$ parts of the ciphertext product, respectively. The ciphertexts that decrypt to the terms of $s^2$ and $s^3$ in the above formula need to be determined. In 2-input ciphertext multiplication, $evk$ is chosen as $(-s\cdot evk_1+e'+Ps^2,evk_1)$. In this case, decrypting $ P^{-1}d_2evk$ leads to $P^{-1}d_2(-s\cdot evk_1+e'+Ps^2+s\cdot evk_1)=P^{-1}d_2(e'+Ps^2)\approx d_2s^2$. Extending this relinearization, the ciphertext that decrypts to $d_3s^3$ can be derived by introducing an extra evaluation key, $evk'=(-s\cdot evk_1+e''+Ps^3,evk_1)$, and multiplying it to $P^{-1}d_3$. Here the same $evk_1$ is utilized for both $evk$ and $evk'$ to reduce the hardware complexity, which will be detailed in the next section. However, a different random vector with low norm, $e''$, is utilized in $evk'$ so that the secret key can not be derived from $evk-evk'$. Accordingly, the final product of three ciphertexts can be calculated as 
\begin{equation}\label{eq: product3}
ct^*=(d_0,d_1)+\lfloor P^{-1}d_2evk+P^{-1}d_3evk'\rceil.
\end{equation}
RNS representation can be also applied to the above formulas. To reduce the noise to a similar level as using two 2-input ciphertext multiplications to compute the product, two rescalings by $q_{L-1}$ and $q_{L-2}$ need to be applied to $ct^*$.

From \eqref{eq: mult3}, 16 2-input polynomial multiplications are required to directly calculate $d_0,d_1,d_2$ and $d_3$. To reduce the polynomial multiplication complexity, the Karatsuba formula is firstly applied to compute $f_0+f_1s+f_2s^2=(c^1_0+c^1_1s)(c^2_0+c^2_1s)$ as 
\begin{equation}\label{eq: f formula}
\begin{split}
    &f_0=c^1_0c^2_0;\  f_2=c^1_1c^2_1\\
    &f_1=(c^1_0+c^1_1)(c^2_0+c^2_1)-f_0-f_2\\
\end{split}
\end{equation}
In the above formulas, $f_0$ and $f_2$ can be shared in the computation of $f_1$. Then $d_0+d_1s+d_2s^2+d_3s^3=(f_0+f_1s+f_2s^2)(c^3_0+c^3_1s)$ remains to be computed. By extending the Karatsuba formula, it can be derived that 
\begin{equation}\label{eq: d formula}
\begin{split}
&d_0=f_0c_0^3;\ d_3=f_2c_1^3;\ g_1=f_1c_1^3; \ g_2=f_2c_0^3\\
&d_1=(f_0+f_1)(c_0^3+c_1^3)-g_1-d_0;\  d_2=g_1+g_2
\end{split}
\end{equation}
Following \eqref{eq: f formula} and \eqref{eq: d formula}, the polynomial multiplication step for 3-input ciphertext multiplication can be completed by 8 multiplications of two polynomials.

\subsection{Noise Analysis}
Associate each ciphertext $ct^t$ with a tuple $(m_t, v_t, B_t)$, where $m_t$ represents the corresponding message polynomial and $v_t$ is the maximum magnitude of the messages. If $ct^t$ is decrypted to $m_t+e_t$, then $B_t$ equals the maximum magnitude of the entries in the error vector $e_t$. 

Decrypting the result of the proposed 3-input ciphertext multiplication leads to 
\begin{equation} \label{eq: 3-ct-dec-noise}
    \begin{split}
        &(m_1+e_1)(m_2+e_2)(m_3+e_3)\\
        &= m_1m_2m_3 + m_1m_2e_3 + m_1e_2m_3 + e_1m_2m_3 \\
        &+ m_1e_2e_3 + e_1m_2e_3 + e_1e_2m_3 + e_1e_2e_3.
    \end{split}
\end{equation}
In the above formula, $m_1m_2m_3$ is the decrypted message while the other terms are noise. Assume that all the three input ciphertexts have the same $B_t=B$ and $v_t=v$. Accordingly, the noise in \eqref{eq: 3-ct-dec-noise} is upper bounded by $3v^2B+3vB^2+B^3$. Denote the noise in $P^{-1}d_2evk$ and $P^{-1}d_3evk'$ by $B_{\text{relin}}$ each. Let the noise caused by rounding be $B_{\text{round}}$. From \eqref{eq: product3} and rescaling, following an analysis similar to that in \cite{CKKS}, the noise in the result of the proposed 3-input ciphertext multiplication is
\begin{equation} \label{eq: noise-3-ct}
    \begin{split}
        B_{3\times} &\leq \frac{1}{q_{L-2}q_{L-1}}(3v^2B\!+\!3vB^2\!+\!B^3 \!+\!2B_{\text{relin}}\!+\!B_{\text{round}}). 
    \end{split}
\end{equation}

Now consider the case that $ct^1$ and $ct^2$ are first multiplied and then the product is multiplied to $ct^3$ using 2-input multiplications. For $ct^1$ and $ct^2$ multiplication, the noise in $(m_1+e_1)(m_2+e_2)$ is bounded by $2vB+B^2$. Considering the relinearization and re-scaling, the noise in the message corresponding to the product of $ct^1$ and $ct^2$, denoted by $B'$, is equal to $\frac{1}{q_{L-1}}(2vB+B^2+B_{\text{relin}}+B_{\text{round}})$. Since two message polynomials are multiplied, the maximum magnitude of the product is $v'=v^2/q_{L-1}$ incorporating the re-scaling. Carrying out a similar analysis on the multiplication of the $ct^1$ and $ct^2$ product and the third ciphertext $ct^3$, it can be derived that the overall noise is 
\begin{equation}\label{eq: noise-2-ct}
    \begin{split}
        &B_{2\times}\leq\frac{1}{q_{L-2}}(v'B+vB'+B'B+B_{\text{relin}}+B_{\text{round}})\\
        &= \frac{1}{q_{L-1}q_{L-2}}(3v^2B+3vB^2+vB_{\text{relin}}+vB_{\text{round}}+B^3\\
        &+BB_{\text{relin}}+BB_{\text{round}}) + \frac{1}{q_{L-2}}(B_{\text{relin}}+B_{\text{round}}).
    \end{split}
\end{equation}

\begin{figure*}[ht]
		\centering
		\includegraphics[scale=0.5]{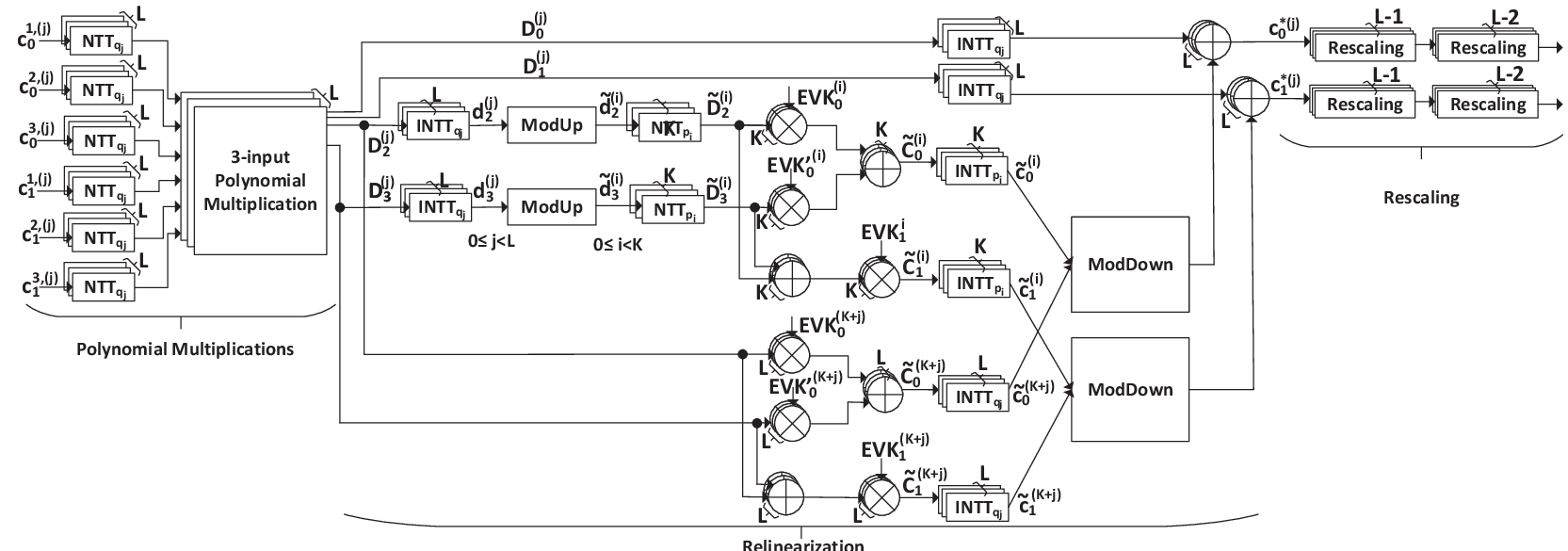}
		\caption{The block diagram for the proposed 3-ciphertext multiplication utilizing the RNS-CKKS scheme.} \label{fig: 3_ct-mult}
\end{figure*}

Comparing \eqref{eq: noise-3-ct} and \eqref{eq: noise-2-ct}, it can be observed that the proposed scheme has lower noise. The reason is that, in the original scheme, the relinearization noise from the first 2-input ciphertext multiplication is multiplied by that of the second 2-input ciphertext multiplication. In contrast, the noises of the two relinearizations are added rather than multiplied in our scheme. 

\section{Three-Input Ciphertext Multiplier Hardware Architecture and Complexity Analysis}
This section presents an efficient hardware architecture for implementing the proposed 3-input ciphertext multiplication. Simplifications are developed to reduce the numbers of ModDown and $evk$ multiplications required. Then the complexity of our proposed design is analyzed and compared with that of prior approach.

The block diagram of the proposed architecture is shown in Fig. \ref{fig: 3_ct-mult}. Similar to the 2-input ciphertext multiplication, all polynomial multiplications are carried out in the NTT domain to reduce the complexity. The multiplications of the polynomials from the three input ciphertexts are implemented according to \eqref{eq: f formula} and \eqref{eq: d formula}. Since ModUp and ModDown are nonlinear operations, INTT is needed before they are carried out. The details of ModUp and ModDown can be found in \cite{RNSCKKS}. From \eqref{eq: product3}, the two polynomials in $P^{-1}d_2evk$ are added to those in $P^{-1}d_2evk'$. If this formula is directly implemented, two ModDown blocks are needed and their outputs are added up. Since the ModDown operations involve modular additions/multiplications with constants, the addition of their outputs can be moved to their inputs and further to before the INTT blocks as shown in Fig. \ref{fig: 3_ct-mult}. As a result, only two ModDown block is needed and the number of INTT blocks at their inputs is reduced to a half. Additionally, our design chooses to use the same $evk_1$ in both evaluation keys, $evk$ and $evk'$. Accordingly, the ModUp results of both $d^{(j)}_2$ and $d^{(j)}_3$ are multiplied to the same $evk_1$ in the relinearization process. Hence, the adders at the outputs of $evk_1$ multiplications can be moved to their inputs. This helps to reduce the number of polynomial multiplications needed. 

\begin{table}[h!]
\begin{center}\caption{Complexities of 2-parallel building blocks in ciphertext multiplications with $L$ and $K$ factors in $Q$ and $P$, respectively.}
\label{tab: modules complexities}
\begin{tabular}{@{}c|@{}c@{}|@{}c@{}|@{}c@{}||@{}c@{}}
\hline
 &
\begin{tabular}{@{}c@{}} 
\# of
\\ 
 Mod. Mult.
\end{tabular} 
 &
\begin{tabular}{@{}c@{}} 
\# of
\\ 
 Mod. Adder
\end{tabular} &
\begin{tabular}{@{}c@{}} 
\# of $w$-bit
\\ 
 data in Memory
\end{tabular}
  & 
\begin{tabular}{@{}c@{}} 
\# of Pipeline.
\\ 
 stages
\end{tabular} \\ \hline
NTT & $\log N$ & $2\log N$  & $N/2\log N$ & $N/2\!-\!1\!+\!5\log N$\\ \hline
INTT & $\log N$ & $4\log N$ & $N/2 \log N$ & $N/2\!-\!1\!+\!5\log N$\\ \hline
ModUp & $2L+2LK$ & $2(L-1)$ & $L+LK$ & $7$\\ \hline
ModDown & $2L\!\!+\!\!2K\!\!+\!\!2LK$ & $2K$ & $K+LK$ & $10$\\ \hline
Rescaling & $2(L-1)$ & $2(L\!-\!1)$ & $L(L\!-\!1)/2$ & $4$ \\ \hline
\begin{tabular}{@{}c@{}} 
2-input poly.
\\ 
 mult.
\end{tabular} 
& $6L$ & $8L$ & $0$ & $4$ \\ \hline
\begin{tabular}{@{}c@{}} 
3-input poly.
\\ 
 mult.
\end{tabular} 
& $16L$ & $18L$& $0$ & $8$ \\ \hline
\end{tabular}
\end{center}
\end{table} 

The 2-parallel NTT and INTT designs in \parencite{ParhiNTT} are adopted in our proposed scheme. Such an NTT or INTT block comprises $\log_2 N$ processing elements (PEs) with delay and switching elements in between. A PE for NTT has one modular multiplier and two modular adders, while that of INTT has one modular multiplier, four modular adders, and two multiplexors. Each of these PEs also has a memory of size $N/2\times w$ for storing twiddle factors. Here $w$ is the bit width of $q_j$. Such a 2-parallel NTT/INTT generates two outputs in each clock cycle. Since the moduli are pre-determined constant, the polynomial coefficient modular multiplier can be implemented utilizing the Barrett reduction with three pipelining stages \parencite{ParhiNTT} and one $w$-bit multiplier in the critical path. Accordingly, the PEs in the NTT/INTT are pipelined into five stages. The complexity of the NTT, INTT, ModUp, ModDown, and rescaling blocks that generate two outputs in each clock cycle are listed in Table \ref{tab: modules complexities}. 

\begin{table}[h!]
\begin{center}\caption{Complexity of the proposed 3-input ciphertext multiplication compared to applying 2-input ciphertext multiplication twice.}
\label{tab: complexity campare}
\begin{tabular}{@{}c|@{}c@{}|@{}c@{}}
\hline
 & 
\begin{tabular}{@{}c@{}} 
2 copies of 2-input
\\ 
 multiplier (Fig. \ref{fig: 2-ct_mult})
\end{tabular}  
& 
\begin{tabular}{@{}c@{}} 
proposed 3-input
\\ 
 multiplier (Fig. \ref{fig: 3_ct-mult})
\end{tabular} \\ \hline
\# of Mod. Mult. 
& 
\begin{tabular}{@{}c@{}} 
$(18L+6K)\log N-8$
\\ 
 $+40L+16K+12LK$
\end{tabular} 
& 
\begin{tabular}{@{}c@{}} 
$(12L+4K)\log N-12$
\\ 
 $+35L+7K+8LK$
\end{tabular} 
\\ \hline
\# of Mod. Adder &
\begin{tabular}{@{}c@{}} 
$(56L+20K)\log N-8$
\\ 
 $+32L+12LK-2K$
\end{tabular} 
& 
\begin{tabular}{@{}c@{}} 
$(36L+12K)\log N$
\\ 
 $+34L+8LK-12$
\end{tabular} 
\\ \hline
\begin{tabular}{@{}c@{}} 
\# of $w$-bit data
\\ 
  in Memory
\end{tabular}
& 
\begin{tabular}{@{}c@{}} 
$(L\!+\!K)N\log N\!+\!L\!+\!K$
\\ 
 $+2LK+L(L-1)/2$
\end{tabular} 
& 
\begin{tabular}{@{}c@{}} 
$(L\!+\!K)N\log N\!+\!L\!+\!K$
\\ 
 $+2LK+L(L-1)/2$
\end{tabular} \\ \hline
\begin{tabular}{@{}c@{}} 
\# of Delay
\\ 
 elements
\end{tabular} 
& 
\begin{tabular}{@{}c@{}} 
$(60L\!+\!12K)(N/2-1$
\\ 
 $+5\log N)\!+\!440L\!+\!40K$
\end{tabular} 
&
\begin{tabular}{@{}c@{}} 
$(40L\!+\!8K)(N/2-1$
\\ 
 $+5\log N)\!+\!386L\!+\!34K$
\end{tabular}
\\ \hline \hline
Latency (clk) & $4N+40\log N+ 46$ & $2N+20\log N + 32$ \\ \hline
\end{tabular}
\end{center}
\end{table}

\vspace{-0.2em}

Adding up the complexities of the building blocks, the overall complexity of our proposed 3-input ciphertext multiplier is summarized in Table \ref{tab: complexity campare}. The design in Fig. \ref{fig: 2-ct_mult} is instantiated twice for a fair comparison and its complexity is also included in this table. Since the memories are used to store the twiddle factors, they can be shared among the NTT and INTT blocks. In the case that the blocks are reading the twiddle factors in the same sequence but at different clock cycles, registers are used to temporarily hold the twiddle factors instead of duplicating the memories. Both of the ciphertext multipliers listed in Table \ref{tab: complexity campare} are fully pipelined. 

A $w$-bit modular multiplier comprises three multipliers, two carry-ripple adders, one comparator, and one multiplexer \parencite{ModMulBarrett}; A modular adder can be implemented by two adders, one comparator, and one multiplexer. A $w$-bit carry-ripple adder consists of $w$ full adders (FAs), while a multiplier requires $w(w-1)$ FAs. Each full adder (FA) can be implemented using an area equivalent to 4.5 XOR gates. The area of each multiplexer is equivalent to that of a single XOR gate. Additionally, the area of 1-bit memory is considered to be equal to that of 1 XOR gate, while a delay element occupies an area equivalent to 3 XOR gates. For an example case of $L=K=3$, $\log_2 N=12$ and $w=30$, our proposed three-input multiplier design leads to $29\%$ smaller area and $50\%$ shorter latency compared to that of two concatenated 2-input multipliers, while having the same critical path. From Table \ref{tab: complexity campare}, it can be observed that the latency improvement achievable by our proposed design does not change with $w$, $L$, $K$, or $N$. Also our design requires much smaller numbers of adders and multipliers regardless of the values of $w$, $L$, $K$, and $N$. Therefore, our design can always achieve significant area saving for various parameters. 

\section{Conclusions}
This paper proposes a three-input ciphertext multiplication for the RNS-CKKS HE scheme by introducing an extra evaluation key. The input polynomial multiplications and relinearization are reformulated to enable the sharing of intermediate results and the expensive modulus switchings. Besides, the evaluation key is chosen to simplify the multiplications. Compared to using two concatenated 2-input multipliers to calculate the product of three ciphertexts, the proposed scheme shortens the latency to almost a half and reduces the area substantially with lower noise. Future work will extend to multipliers with more than three inputs.

\printbibliography

\end{document}